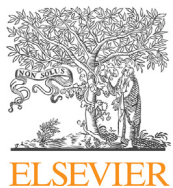

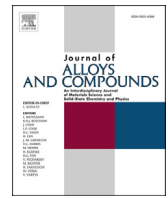

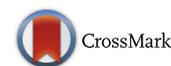

# Thermal evolution of nanocrystalline co-sputtered Ni−Zr alloy films: Structural, magnetic and MD simulation studies

Debarati Bhattacharya [a, *], T.V. Chandrasekhar Rao [b], K.G. Bhushan [b], Kawsar Ali [c], A. Debnath [b], S. Singh [a], A. Arya [c], S. Bhattacharya [b], S. Basu [a]

[a] Solid State Physics Division, Bhabha Atomic Research Centre, Mumbai 400085, India
[b] Technical Physics Division, Bhabha Atomic Research Centre, Mumbai 400085, India
[c] Material Science Division, Bhabha Atomic Research Centre, Mumbai 400085, India

## ABSTRACT

Monophasic and homogeneous $Ni_{10}Zr_7$ nanocrystalline alloy films were successfully grown at room temperature by co-sputtering in an indigenously developed three-gun DC/RF magnetron sputtering unit. The films could be produced with long-range crystallographic and chemical order in the alloy, thus overcoming the widely acknowledged inherent proclivity of the glass forming Ni−Zr couple towards amorphization. Crystallinity of these alloys is a desirable feature with regard to improved efficacy in applications such as hydrogen storage, catalytic activity and nuclear reactor engineering, to name a few. Thermal stability of this crystalline phase, being vital for transition to viable applications, was investigated through systematic annealing of the alloy films at 473 K, 673 K and 923 K for various durations. While the films were stable at 473 K, the effect of annealing at 673 K was to create segregation into nanocrystalline Ni (superparamagnetic) and amorphous Ni + Zr (non-magnetic) phases. Detailed analyses of the physical and magnetic structures before and after annealing were performed through several techniques effectual in analyzing stratified configurations and the findings were all consistent with each other. Polarized neutron and X-ray reflectometry, grazing incidence x-ray diffraction, time-of-flight secondary ion mass spectroscopy and X-ray photoelectron spectroscopy were used to gauge phase separation at nanometer length scales. SQUID based magnetometry was used to investigate macroscopic magnetic properties. Simulated annealing performed on this system using molecular dynamic calculations corroborated well with the experimental results. This study provides a thorough understanding of the creation and thermal evolution of a crystalline Ni−Zr alloy.

## 1. Introduction

Ni−Zr alloys are well known to have high glass-forming ability and are thus popular as bulk metallic glasses. Amorphous Ni−Zr alloys formed by solid state reaction are stable over a wide composition range [1], consequently presenting a high nucleation barrier to the crystalline phase. The multitude of existing research on amorphous Ni−Zr alloys also bears testimony to this premise. Bulk Ni−Zr alloys are prepared through liquid quenching, which being a quick process, does not allow sufficient time for reorganization of atoms to form crystalline phases, thus leading to amorphous structures [2]. But they can be crystallized through subsequent thermal treatments [3]. On the other hand, Ni−Zr alloy films have been prepared by amorphization of crystalline multilayers [4], diffusion couples [5] or crystallization of amorphous alloys [6]. Studies on Ni−Zr alloy films are mainly devoted to amorphous structures along with their widespread use in varied fields as in catalytic activity and for some unique applications like fuel cells [7], in biomedical industry [8], as stressed atomic force microscopy probes [9] and extensively in nuclear reactor engineering and hydrogen storage [10] and separation [11]. However, the formation of crystalline films of these alloys directly through some means has not been focused upon earlier. The feature of crystallinity can promote enhanced utility in the numerous applications associated with amorphous Ni−Zr alloys. For example, nanocrystallinity would be desirable for hydrogen storage materials since it could promote better hydrogen sorption properties [12]. But this also requires the crystalline films to be

---

thermodynamically stable. The present study aims at producing nanocrystalline Ni–Zr alloy films by magnetron sputtering and studying thermodynamic stability of the crystalline phase. The proclivity of the Ni–Zr system towards amorphization could be overcome by depositing thin films of $Ni_{10}Zr_7$ alloy in a controlled manner by magnetron sputtering. Apart from the known advantages of this technique like large-area film deposition with reproducible characteristics, sputtered atoms are energetic enough to produce large number of nucleation sites. The growth of these nucleation sites can be slowed down through controlled deposition rate, ensuring the onset of crystallinity. A very effective method of generating binary alloy films is that of simultaneous sputtering (co-sputtering) of two elements. Sputter rates of individual targets can be controlled leading to uniformity of alloys prepared in chosen stoichiometry. Co-sputtering can also lead to amorphous films as seen in the case of Ni–Zr system [13] and [14]. It is well known that choice of deposition parameters of the sputtering process directly affects the structure and properties of the films deposited [15] and [16]. In general, by controlling deposition parameters such as sputter power and thereby deposition rate [17] and [18], residual gas pressure and target–substrate distance, it is possible to obtain crystalline alloys through co-sputtering.

There exist many reported studies on the effect of annealing on alloys of various materials to create phase separation which can lead to more interesting structures. Studies pertaining to alloy element segregation due to annealing are important and have been employed for several systems like catalysts [19] and widely for core shell structures [20,21], and [22]. Some applications of crystalline magnetic species embedded in an amorphous matrix formed through structural transformation of annealed alloys have also been reported [23]. Segregation of alloying elements to surface and grain boundaries is vital for the study of structural stability of materials subjected to prolonged exposure at elevated temperatures in a reactor environment [24]. Ni–Zr alloys are important for their possible application in nuclear industry in various parts of coolant flow loops. Thermodynamic stability of such alloys under heat treatment is of paramount importance regarding various aspects of their utility viz. life span, usable temperature range and mechanism of heat transport. Phase stratification to produce different phases with annealing has been studied in amorphous Ni–Zr alloys [25]. Surface segregation in Zr alloys is also known to be caused by oxygen content within a sample or adsorbed on the surface [26] and [27]. Most of the simulation work reported on Ni–Zr systems is related to their amorphization behavior using molecular dynamics (MD) method [28,29,30,31], and [32]. Earlier, Li et al. [28] studied the change in volume on glass transition in $NiZr_2$ using MD simulations. Delogu [30] studied the effect of lattice distortions on the glass-forming ability in Ni–Zr system. Massobrio et al. [31] used MD simulation to show the effect of chemical disorder (antisite defects) on the formation of amorphous phase from crystalline $NiZr_2$. The diffusion coefficients of Ni and Zr and its relation to the glass transition temperature in Ni20Zr80 glassy phase have been determined by Mutiara et al. using MD simulations with Stirling-Weber potential [32]. To the best of our knowledge, simulation studies on the segregation behavior in the amorphous phase have not been reported earlier.

This paper reports the successful formation of nanocrystalline alloy films of $Ni_{10}Zr_7$ by simultaneously sputtering Ni and Zr targets at very slow rates in an indigenously built DC/RF magnetron sputtering system. The $Ni_{10}Zr_7$ phase is a stable one according to the Ni–Zr phase diagram, with the second lowest free energy associated with its formation; the first being equiatomic NiZr phase. Hence this stoichiometry was chosen in order to facilitate the study of thermal stability of the nanocrystalline state. These alloy films were then annealed at 473 K, 673 K and 923 K under vacuum for fixed periods. Several surface and interface sensitive techniques were used to characterize the as-deposited and annealed films for their physical and magnetic structure. X-ray reflectivity (XRR), polarized neutron reflectivity (PNR) and time-of-flight secondary ion mass spectrometry (TOF-SIMS) were used for analyses of structural changes between as-deposited and annealed Ni–Zr alloys. Grazing incidence x-ray diffraction (GIXRD) performed at 0.5° incidence could identify crystalline peaks observed before and after annealing. The $Ni_{10}Zr_7$ phase being the most stable among the Ni-rich Ni–Zr alloys enabled the investigation of magnetic properties induced due to annealing. Magnetic characterization of these samples was achieved using both PNR and SQUID-based magnetometry. X-ray photoelectron spectroscopy (XPS) was also used to obtain elemental profiles on the surface before and after annealing of the samples. Moreover, MD simulations were employed to determine the clustering tendencies present after annealing, using bulk and surface slab models. The migration energies of Ni and Zr atoms in the as-deposited case were also determined using MD calculations.

## 2. Experimental

### 2.1. Deposition and characterization of $Ni_{10}Zr_7$ alloy films

The Ni–Zr alloys used for this study were deposited by co-sputtering in a specially designed and developed 3-gun DC/RF magnetron sputter system [33]. Several calibration experiments were carried out prior to deposition in order to determine the variation of thickness with factors affecting deposition rate of each target material: the power applied to the source, source-substrate distance and sputter gas (Ar) pressure. These experiments were carried out for each target since sputter rates are essentially material dependent. Film thicknesses were obtained through XRR data fitting. The Ni and Zr targets used had a high purity of ~99.95% and were pre-sputtered before use. Two inch diameter Si wafers were used as substrates and held at 85 mm from the target. Adequate measures were adopted for all procedures to maintain a clean environment so that resulting samples could be made impurity-free. Substrates were chemically cleaned and mounted in the vacuum chamber through a load lock system, to minimize contamination and maintain ultimate vacuum of ~$10^{-7}$ Torr. The argon background gas flow was tuned to 20 sccm through a mass flow controller, regulating pressure of 3.75 mTorr during deposition. The gas pressure is an important parameter since multiple collisions with the background gas can alter energy distribution of atoms and hence their sticking probability. Homogenous and uniform alloy films were formed on the substrates by co-sputtering Ni and Zr targets and maintaining very slow deposition rates as Ni: 0.4 Å/sec and Zr: 0.1 Å/sec through careful D.C. power control. Different atomic ratios of Ni/Zr could be obtained by varying pulsed D.C. power settings for the two sputtering targets. The corresponding compositions of the samples studied were determined using X-ray and polarized neutron reflectivity data for the same samples. The final composition of $Ni_{10}Zr_7$ had atomic ratio of Ni/Zr as 1.4; indicated in Fig. 1 along with the relevant sputter powers employed.

The alloy films were progressively annealed at 473 K, 673 K and 923 K in another vacuum chamber for varying time periods: 2 h, 4 h and 8 h. After each anneal, the films were characterized through the techniques described below.

XRR and PNR are used as complementary tools to probe chemical and magnetic profiles of layered structures. The high degree of depth sensitivity of these powerful techniques to differences in refractive indices of adjacent layers essentially enables structural characterization in a stratified structure with spatial resolution in fractions of 1 nm. Specularly reflected X-ray and neutron intensities at grazing incidence angle $\theta$ on the sample surface were recorded as a function of momentum transfer $q = (4\pi/\lambda)\sin\theta$ in $Å^{-1}$ (where $\lambda$ is the wavelength of incident beam), normal to the reflecting surface. Further, for PNR, the spin of the neutron beam was set parallel (spin-up) or anti-parallel (spin-down) to the in-plane sample magnetization. The spin-up reflectivity and the spin-down reflectivity are termed $R^+$ and $R^-$ respectively. Refractive index for polarized neutrons depends on the relative orientation of the sample magnetization and the neutron spin. The PNR data yields depth dependent magnetic profile in a sample. XRR and PNR data were independently analyzed with a model of depth-dependent scattering length density (SLD) averaged over lateral dimensions of the entire sample. SLD is a function of number density of the atomic components and has units of $Å^{-2}$. For x-rays, this corresponds to *electronic* SLD (ESLD). In case of PNR, they correspond to coherent *nuclear* SLD (NSLD) and *magnetic* SLD (MSLD), which is the magnetization profile as a function of depth in the sample. It was from this generated structure that quantitative estimates of thickness, roughness, density of the relevant layers and magnetic moment density profile with excellent depth resolution for the samples used in the present study, were derived. The typical error in each of the model parameters was estimated using Bootstrap method for error analysis. Errors on the parameters obtained from the fits were found to be within 5%. Atomic number densities of the constituents of each layer could also be extracted and thus the joint analysis method using XRR and PNR together provides an excellent handle to obtain stoichiometry of binary alloy layers. Details with regard to this procedure as well as the formalism used for data fitting can be found in an earlier work [34].

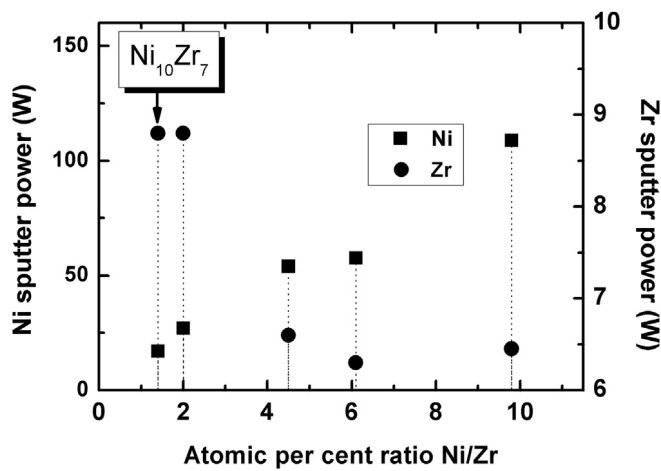

**Fig. 1.** Composition of Ni–Zr alloys controlled through variation of sputter power coupled to target sources. Dotted lines joining pairs of Ni and Zr sputter powers relate to various Ni/Zr ratios attempted. Final composition of $Ni_{10}Zr_7$ is indicated for a specific such pair.

The XRR and GIXRD measurements were both made in parallel beam geometry with a powder diffractometer using Cu $K_\alpha$ radiation. It was ensured that data was recorded from the entire sample since the X-ray beam in this mode falls completely on the entire sample, unlike in Bragg-Brentano mode wherein the X-ray spot size versus sample dimension (geometry) requires correction or scaling. PNR data were recorded on the reflectometer set up at DHRUVA reactor, India [35]. TOF-SIMS was used for obtaining elemental profiles of the samples as a function of depth. It was performed using a set up developed in-house with a 25 KeV $^{69}$Ga mono-isotopic liquid metal ion gun as the primary ion source giving a spot size of about 250 nm. Measurements to determine the magnetic nature of the films were carried out using a SQUID magnetometer (Quantum Design, model MPMS), with field applied parallel to the sample surface. Samples were sandwiched between pieces of straw and no separate sample holder was used.

### 2.2. Details of MD simulations used

The MD simulations as embedded in LAMMPS (Large-Scale Atomic/Molecular Massively Parallel Simulator) code (http://lammps.sandia.gov) [36] were carried out using Finnis-Sinclair embedded-atom method (EAM-FS) [37] potential as developed by Mendelev [38]. In EAM-FS, the total energy $E_i$ of an atom '$i$' is given by $[F_\alpha\{\sum\rho_{\alpha\beta}(r_{ij})\} + (1/2)\{\sum\varphi_{\alpha\beta}(r_{ij})\}]$, where '$i$' and '$j$' are sites at distances '$r$' occupied by $\alpha$ and $\beta$ type of atoms, $\varphi_{\alpha\beta}(r_{ij})$ is the two body pairwise potential between atoms $i$ and $j$ separated by a distance $r_{ij}$ and $F_\alpha(\rho)$ is the embedding energy which depends upon the electron density, $\rho_{\alpha\beta}$. The latter is a function specific to the atomic types of both atoms $i$ and $j$, making a distinction between different elements which can contribute to the total charge density at a given site depending upon the identity of the atom at that site.

Both bulk and surface slab calculations were carried out to study segregation effects upon annealing in $Ni_{10}Zr_7$ alloy. For bulk calculations, a simulation box of $5 \times 5 \times 5$ supercell of original orthorhombic unit cell was employed which contained 8500 (5000 Ni and 3500 Zr) atoms with periodic boundary conditions in the three directions (i.e. no surface effect). A NPT (constant N: number of particles, P: pressure and T: temperature) ensemble with Nose-Hoover thermostat and barostat was used to control the temperature at a pressure of 1 atm. For surface calculations, a simulation box of length 5 unit cells in X and Y directions and 20 unit cells along Z direction was chosen, which contained 34,000 (20,000 Ni and 14,000 Zr) atoms with periodic boundary conditions in X- and Y-directions and fixed boundary along Z-direction. In this case the temperature was controlled by Langevin thermostat and pressure was controlled by Berendsen barostat. For both bulk and surface slab, the simulated annealing was carried out at 673 K and 873 K for a period of 200 picoseconds each and nearest neighbor analysis around both Ni and Zr atoms over a certain cut-off distance was performed before and after the annealing.

Through MD simulations diffusion barriers for both Ni and Zr atoms in the lattice using nudged-elastic-band (NEB) method [39,40], and [41] as embedded in LAMMPS were determined. In the vacancy diffusion mechanism, a vacancy was created at the nearest neighbor of the diffusion atom in the simulation box in the primary state of diffusion. In the final state, the diffused atom arrived in the vacancy position. There were 48 intermediate states inserted between these two states. The minimum energy path (MEP) was determined and the diffusion barrier was calculated in the MEP.

## 3. Results and discussion

### 3.1. Structural characterization

XRR patterns of the as-deposited and 473 K, 673 K and 923 K annealed (@ 2 h, 4 h and 8 h respectively) alloy films were recorded and structural parameters were extracted through suitable model fits to the data as explained in the previous section. There was no significant change observed for the 473 K annealed films; but after annealing at 673 K for 2 h, the reflectivity patterns appeared very different. The reflectivity data and corresponding fits (solid lines) of as-deposited and only 2 h annealed (at 673 K) films are shown for comparison in Fig. 2a. The three ESLD depth profiles of annealed films (673 K for 2 h, 4 h and 8 h) along with as-deposited one are compared in Fig. 2b.

From the widely differing profiles in Fig. 2 it is apparent that after the first anneal at 673 K, an obvious change has been induced

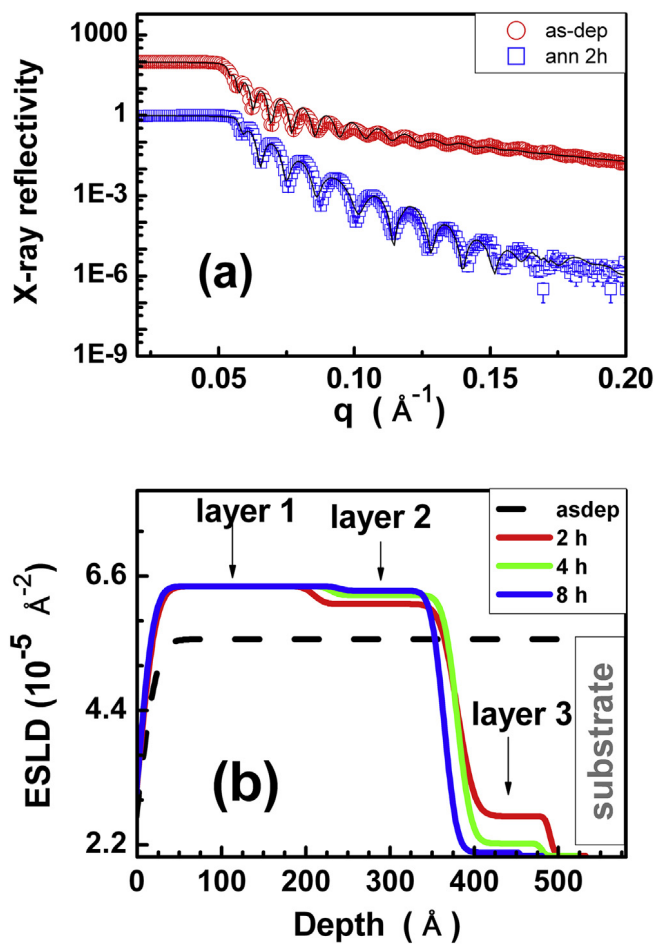

**Fig. 2.** (a) XRR data (open symbols) of as-deposited and 673 K annealed (2 h) along with corresponding fits (solid lines). (b) Electron scattering length density profile of as-deposited and 673 K annealed (2 h, 4 h, 8 h) alloy films. The layered structure formed after annealing can be clearly distinguished. The position of the substrate is indicated only for as-deposited film.

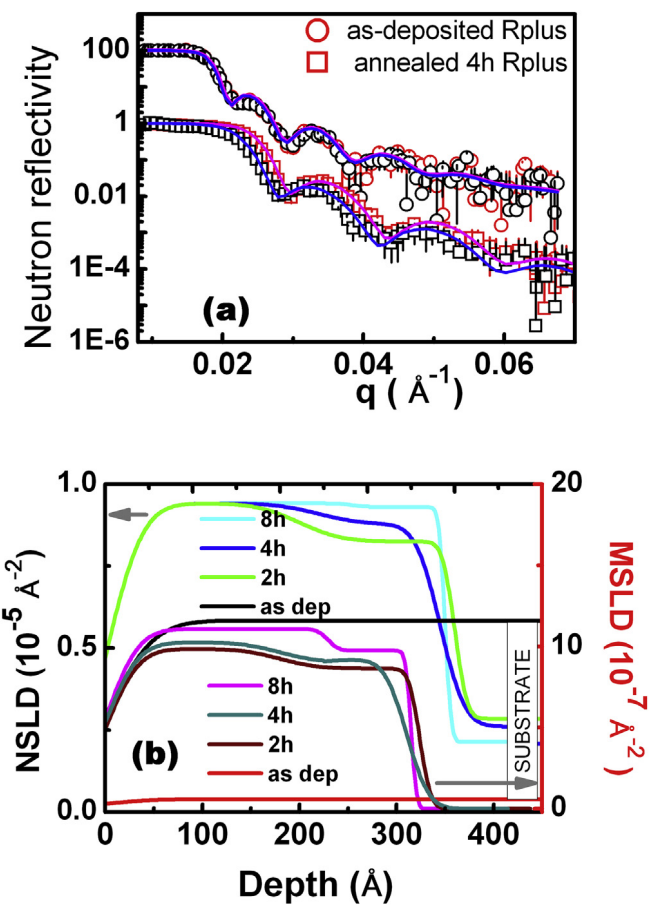

**Fig. 3.** (a) PNR data of as-deposited (open circles) and 673 K annealed at 4 h (open squares) along with corresponding fits (solid lines). The red and black symbols denote spin-up and spin-down data respectively (b) Nuclear and magnetic scattering length density profiles of as-deposited and 673 K annealed (2 h, 4 h, 8 h) Ni–Zr alloy films. Grey arrows indicate the corresponding ordinate axes. Substrate position is indicated only for as-deposited film. (For interpretation of the references to colour in this figure legend, the reader is referred to the web version of this article.)

in the overall structure of the alloy film. The as-deposited alloy data was fitted with a single layer of thickness ~545 Å and the ESLD value corresponded exactly to that of the phase Ni$_{10}$Zr$_7$. The film was also determined to be homogenous and uniform, by scrutinizing various zones with XRR using fine slits, and the data recorded were all the same. However, XRR data of the annealed alloy could be fitted best only on consideration of 3 layers with varying thickness and composition. This layered structure is seen marked in the derived ESLD plots in Fig. 2(b). The thickest top layer of ~200 Å consisted almost entirely of Ni while the lower layers (towards the substrate) had progressively decreasing thickness as well as Ni content. The total sample thickness after annealing appeared to have decreased to approximately 13% of the as-deposited sample. The PNR data of as-deposited and annealed samples were fitted with similar models used for respective XRR analysis. Fig. 3a compares PNR data of spin-up ($R^+$) and spin-down ($R^-$) neutrons along with equivalent fits for as-deposited and only 4 h, 673 K annealed films. The fitted NSLD and MSLD profiles of as-deposited and of films annealed at 673 K for 2 h, 4 h, and 8 h are all represented in Fig. 3b. The features of this figure will be elucidated in Section 3.2. Using the fits to XRR and PNR data and applying the formalism used by Bhattacharya et al. [34] for the extraction of atomic number density profiles and subsequent estimation of composition, it was found that the as-deposited film indeed had a

Ni/Zr atomic ratio as 1.4. This corresponds exactly to the Ni$_{10}$Zr$_7$ phase.

Neither XRR nor PNR are sensitive to differences in crystalline or amorphous nature of the samples studied. The crystallographic structures of the as-deposited and annealed samples were determined by GIXRD measurements. Confirmation of an orthorhombic phase for as-deposited film was obtained through GIXRD data analysis (indexing and fitting with freely available powder crystal structure program), wherein the crystalline peaks obtained could be identified with Ni$_{10}$Zr$_7$ structure (JC-PCPDF 04-007-2407) and the pattern was fitted with an R-factor of 0.005. No peaks corresponding to any other phase were obtained hence monophasic formation of Ni$_{10}$Zr$_7$ was inferred. GIXRD of as-deposited and 8 h annealed alloy films are overlaid in Fig. 4. The annealed film exhibits sharply crystalline peaks corresponding to fcc Ni. This matches the result of reflectivity analyses where the top layer comprises 98.95% Ni.

The results of PNR, XRR and GIXRD clearly indicated that a phase separation of crystalline Ni phase from the alloy phase had occurred near the film-air interface. The Ni–Zr content in the rest of the annealed film existed in non-crystalline state. From the GIXRD patterns, grain sizes of films before and after annealing (673 K) were estimated using Scherrer equation. These values have been plotted against annealing time in Fig. 5, with the star symbol representing average grain size of the as-deposited alloy film and the square symbols being the corresponding values for Ni grains obtained after annealing. The same figure also shows the rise in thickness of top layer for successive annealing times and concomitant shrinkage of overall thickness of annealed films. Grain sizes of as-deposited alloy crystallites were found to be very small (~4 nm) and after the first anneal, the Ni grains were 2–3 times this size. Further annealing times only caused very slow rise in Ni grain growth.

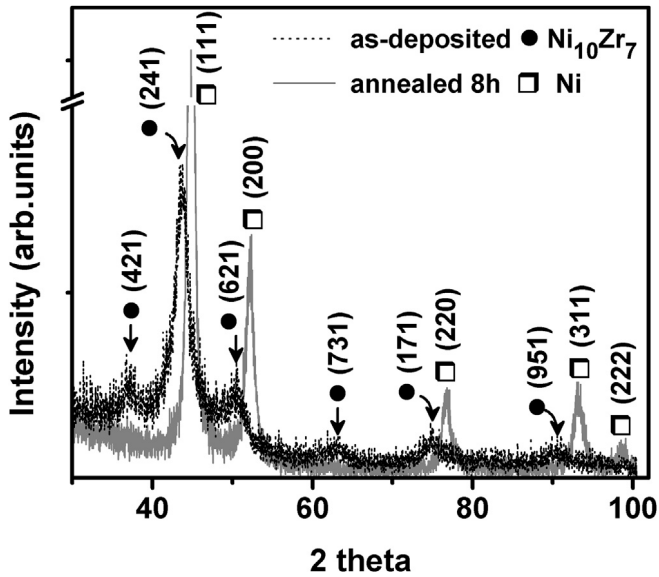

**Fig. 4.** GIXRD (at 0.5°) of as-deposited and annealed Ni–Zr alloy films.

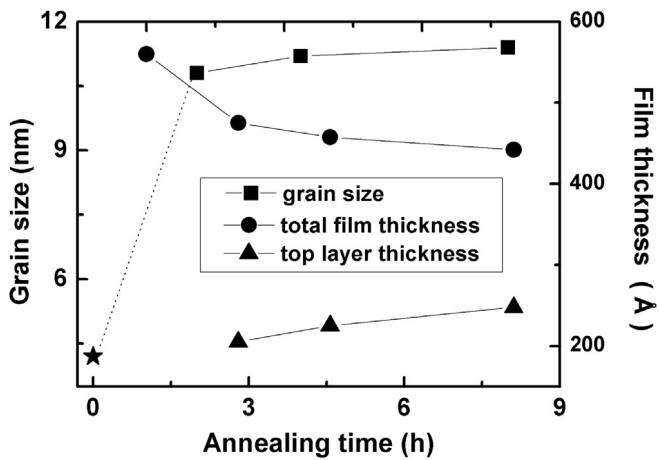

**Fig. 5.** Variation of average grain sizes of Ni–Zr alloy films annealed at 673 K with time, starting from as-deposited film (star). Connecting lines merely serve as visual guides. As a result of annealing, the top layer thickness increases and total film thickness decreases.

A brief discussion on the formation of co-sputtered crystalline $Ni_{10}Zr_7$ alloy in the context of Ni–Zr binary system seems pertinent here. According to the phase diagram of this system, the lowest free energy is associated with NiZr phase and the next most favorable alloy composition in terms of highly negative enthalpy of formation (−48 kJ/mol) is that of $Ni_{10}Zr_7$ formed by peritectic reaction. There appears to be a composition range over which this alloy is achievable as seen in the Ni–Zr phase diagram [42]. This allows for a certain degree of flexibility with regard to variation in sputter power and consequently the deposition rate, in order to adjust the composition. According to Hume-Rothery principle, atomic size difference of greater than 15% between elements of a binary alloy usually leads to limited solid solubility. In case of Ni–Zr system, there exists a large difference between atomic radii of the two components. It is also known that solid solubility of Zr in Ni is poor (<1 at %) and that of Ni in Zr is further reduced [43]. However thermodynamic interaction is governed by heat of mixing and the large negative enthalpy value associated with this system would point to strong bonding interaction between Ni and Zr. This serves as the driving force to produce amorphous structures, which is commonly observed. The high cohesive energies associated with Ni–Zr binary alloys also reflect their structural and mechanical stability [44]. During the sputtering process, rapid creation of nucleation sites can be controlled through deposition rate thus ensuring the onset of crystallinity. Further growth of crystallites is then tuned to follow an ordered structure. This feature has been exploited in this work to form crystalline $Ni_{10}Zr_7$ alloy films by co-sputtering. Low deposition rates do not unduly affect gas temperature either, and thus do not disturb the thermalization process [15] necessary for uniform deposition of the films.

After the first 2 h annealing at 673 K, the alloy film separated into 3 distinct layers as estimated through reflectivity measurements (SLD plots of Figs. 2b and 3b). With further annealing these layers exhibited changes in both thickness and density (composition). The atomic number densities of Ni and Zr corresponding to each layer were calculated through joint XRR and PNR data analyses referred to earlier. These values of atoms/cc are plotted against depth for 2 h and 4 h anneals (at 673 K) in Fig. 6a, clearly showing the 3-layer structure. Here the inherent roughnesses of each surface and interface, also obtained from reflectivity analyses, have been ignored and the plots thus have sharp steps. The 3 layers are indicated and the depth axis proceeding from left to right is correspondingly designated as I: first layer on top (air-film interface), II: intermediate layer and III: last layer before the substrate. The number densities of Ni appeared to drop with annealing time across the layers from the air-film interface to the substrate, accompanied by a slow rise in Zr atoms/cc. The first layer grew in size with longer times of anneal (also evident in Fig. 6a) at the expense of the other layers. Since this was determined to be predominantly composed of Ni, the growth of this layer can be correlated with the slow increase observed in grain size with annealing time. Further annealing at 923 K did not cause any significant modifications in structure.

The TOF-SIMS data for Ni and Zr recorded for the 2 h annealed alloy film are displayed as a depth profile in Fig. 6b. The concentration profile was obtained from the recorded mass spectrum by fitting selected mass peaks and integrating the area under the curve to yield the individual elemental/isotopic concentrations. This profile shows very similar behavior of Ni and Zr concentrations along the depth of the film as obtained from reflectivity analyses, thus testifying to the accuracy of this powerful technique for investigating layered structures. The insets in Fig. 6b include an image of the layer-by-layer depths sputtered (etched) from the film and a plot of etch profile, which shows the length dimension of etched area. TOF-SIMS was also applied to as-deposited film (data not shown here) and by analyzing different depths removed by etching; it was found that the atomic concentrations were constant throughout the film, which reconfirms the compositional uniformity of the as-deposited film with depth. It was also found from analyses of the data that both as-deposited and annealed films contained pure Ni and Zr components and were essentially free from impurities, as decided by the detection limit of 10 ppm of this sensitive technique. The top few surface layers (~200 Å) of as-deposited and 2 h annealed films were also probed by x-ray photoelectron spectroscopy. It is obvious from the scans shown in Fig. 7 that while the as-deposited film (a) bears presence of Zr on the surface, after annealing (b) there is almost no signature of the same. Both samples did show Ni peaks though (scans not shown here); which also points to the segregation of Ni to the top surface. In addition the absence of physisorbed oxygen in XPS scans implied that the surface was free from metal oxide. Hence the segregation cannot be attributed to adsorbed oxygen either. The insets in Fig. 7a and b are energy-dispersive x-ray mapping images of Zr for the respective films. The green component signifies the presence of Zr and the black background constitutes its absence. The inset of Fig. 7a not only shows larger concentration of Zr for as-deposited case but also testifies to the uniform distribution of atomic constituents (mapping of Ni content is not shown here) and homogeneity of the film. The inset in Fig. 7b shows very sparse distribution of Zr on the surface after annealing. Thus Ni has indeed segregated to the top after annealing as corroborated by reflectivity analyses and also evinced by MD simulations. It may be noted here that the surface free energies of Ni and Zr are not so widely different [45].

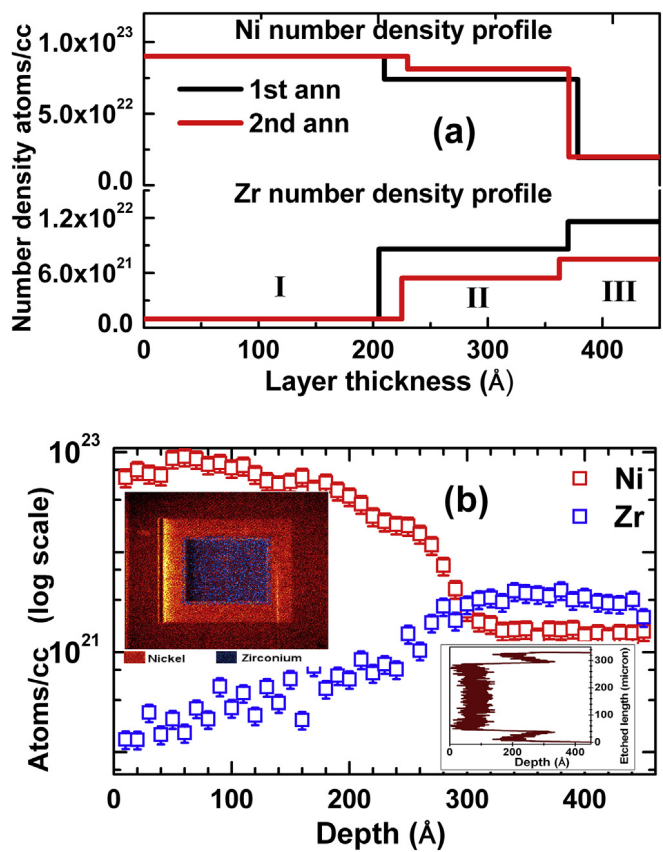

**Fig. 6.** (a) Atomic concentrations of Ni and Zr calculated from reflectivity analyses along the depth of annealed films. The numbered panels are meant to distinguish the layers formed after annealing. Top layer (I), intermediate (II) and last layer on substrate (III). (b) TOF-SIMS profile of Ni and Zr for 2 h annealed film. Left inset: image of etched film at various depths; innermost blue square (lowest layer) contains more of Zr. Right inset: plot of etching area. (For interpretation of the references to colour in this figure legend, the reader is referred to the web version of this article.)

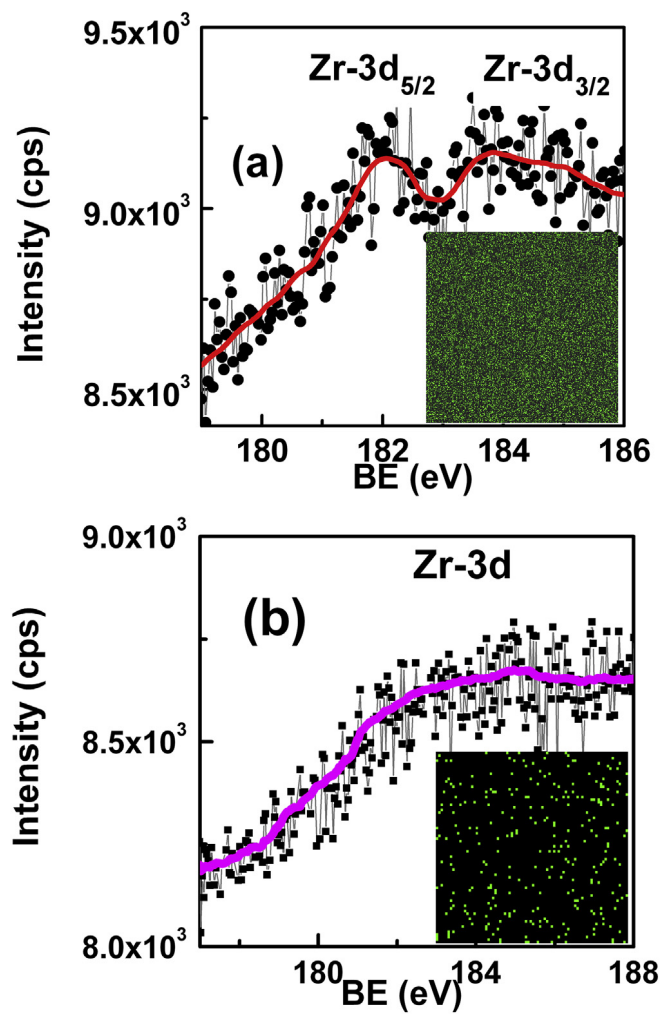

**Fig. 7.** X-ray photoelectron spectroscopy of (a) as-deposited and (b) 2 h annealed alloy film. Insets are energy dispersive x-ray mapping of respective films. The green component is a quantitative measure of Zr content in the 200 Å depth probed. (For interpretation of the references to colour in this figure legend, the reader is referred to the web version of this article.)

The splitting of annealed film into different layers of varying Ni and Zr composition can be explained in terms of the difference in rates of diffusion of the two components. The diffusion of Ni in Zr and also in Ni–Zr alloys is reported to be many times larger than that of the Zr diffusion, regardless of amorphous or crystalline nature of the matrix [46] and [47]. The ratio of sizes of constituent atoms, mentioned earlier, also plays a role in determining the kinetics of diffusion [48]. The relative size effect is the outcome of geometry of the lattice being packed with atoms of different sizes. The smaller Ni atoms can move and be accommodated in the voids around the larger Zr atoms apart from original lattice sites. Movement of atoms within such a structure is accompanied by variations of internal energy, volume and entropy. The diffusional asymmetry mainly leads to disturbance in internal equilibrium of resultant structures [49]. However, rapid diffusion leads to rise in entropy, given the random redistribution of Ni in the system during annealing. When the system is first exposed to rise in thermal energy, the increased internal energy is more than offset by the increased entropy causing the free energy to drop. This is why initial experiments performed with annealing at 473 K (data not shown here) did not show any significant structural changes like phase stratification. However as annealing progresses (at 673 K), the high mobility of Ni causes its saturation and the internal energy developed becomes intolerably high. The system is then driven towards a more stable configuration of reduced free energy by segregating Ni as a separate phase, with the overall structure being dominated by the crystalline Ni lattice. That is why GIXRD shows sharp crystalline peaks corresponding to Ni phase only. The lower layers with greater percentage of Zr were thus construed to be amorphous.

### 3.2. Magnetic measurements

PNR serves as an invaluable tool to determine the magnetization profile, apart from nuclear structure profile, at surfaces and interfaces of thin films and multilayers. A major advantage is that it can detect small magnetic moments with excellent depth resolution. In this technique, the incident neutrons are polarized either parallel (spin-up) or anti-parallel (spin-down) with respect to the magnetized sample, giving rise to two respective components $R^+$ and $R^-$ of the measured reflectivity $R$. The magnetization is in the plane of the film and thus perpendicular to the momentum transfer vector. The difference between the $R^+$ and $R^-$ patterns for each film, as shown in Fig. 3a, is a measure of the magnetic moment density profile associated with that film. Magnetic moment of the as-deposited $Ni_{10}Zr_7$ alloy film was derived from the fitting parameters as ~0.15 $\mu_B$/atom, which is much lower than bulk Ni magnetic moment of ~0.54 $\mu_B$/atom in the ferromagnetic state. After annealing, the magnetic moment rose to ~0.45 $\mu_B$/atom, which approaches the value of bulk Ni and is another evidence of Ni being segregated. The lowermost layer did not contribute to the magnetization of the sample having zero magnetic moment density and was thus concluded to be non-magnetic. Fig. 3b compares the magnetic scattering length densities of all films annealed at 673 K with that of as-deposited one. The drastic difference between magnetic behavior of as-deposited and annealed films explained above is visually obvious here too.

The temperature window for magnetometry was well below the Curie temperature of bulk Ni (627 K). Magnetization measurements were performed on as-deposited and annealed alloy films as a function of temperature and magnetic field.

#### 3.2.1. Temperature dependence of magnetization

Magnetization was measured as a function of temperature (M(T)) under zero-field-cooled (ZFC) and field-cooled (FC) conditions. In the former case, the sample was first cooled to low temperature (5 K) without applying any external field. Once the sample temperature was stabilized, a magnetic field of 100 Oe was switched on and measurements were carried out at various temperatures as the sample warmed up. Fig. 8 shows the results of M(T) measurements on as-deposited and annealed films. While as deposited film had negligible magnetic moment, as expected of non-magnetic $Ni_{10}Zr_7$ alloy, the annealed samples showed significant magnetization on account of pure and nearly-pure Ni phases getting segregated as separate layers. In the absence of an aligning magnetic field during cooling, magnetization of individual particles/domains in the as-deposited film would have remained "frozen" in random directions due to magneto-crystalline anisotropy [50]. Furthermore, since the measuring field applied at 5 K is too low for moments to overcome the anisotropy barrier and align along the field, sample magnetization remained low. As the sample warmed up under 100 Oe, thermal energy provided the impetus for particles with low energy barrier to align their magnetic moments with the field, thereby progressively increasing sample magnetization. The temperature at which energy barriers associated with all the particles are overcome is termed as the blocking temperature.

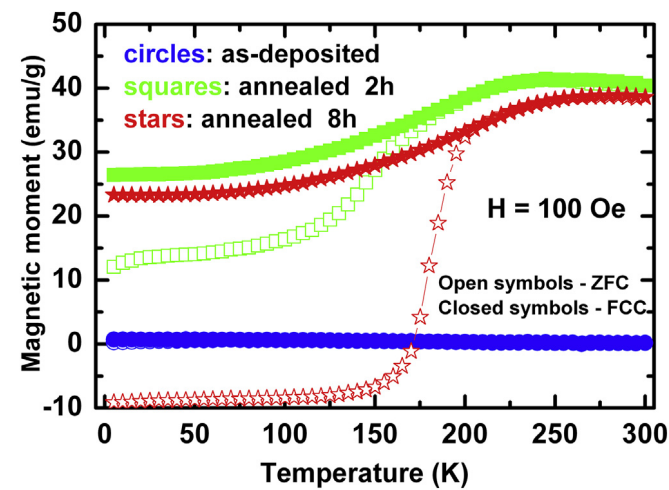

**Fig. 8.** Magnetization of as-deposited and annealed (2 h and 8 h) films as a function of temperature. Measurements were carried out under an applied magnetic field of 100 Oe. The branching between ZFC and FC curves indicates blocking temperatures in annealed cases.

The nature of M(T) is different under FC mode, wherein the sample is cooled under an applied magnetic field from ambient temperature to 5 K and its magnetization measured simultaneously at various temperatures. In this case, moments that are aligned along field direction at ambient temperatures will stay aligned, until some of those would overcome a weak aligning field to disorient on account of significant anisotropy energy difference. Thus, M(T) curve in this case branches-off from ZFC behavior at the blocking temperature and remains nearly temperature independent at lower temperatures but for a small reduction. One can note the branching of ZFC and FC curves at the blocking temperature in case of annealed samples. Their FC curves showed nearly equal magnetization and similar temperature dependence, indicating that the magnetic phases are identical in the two annealed samples. On the contrary, the ZFC behavior of 8 h annealed sample differs distinctly from that of 2 h annealed sample. The former showed a diamagnetic signal till 125 K or so, presumably from Si substrate, before the contribution from Ni layers sets in. This observation leads to the inference that additional annealing has enhanced the particle/magnetic domain size, thus sharpening the distribution of anisotropic energies. Consequently, the blocking temperature rose to 210 K for 8 h annealed sample from 190 K, noted for 2 h annealed sample. Moreover, longer annealing appeared to have completely randomized magnetization of individual Ni particles/domains, negating their contribution to sample magnetization at temperatures below ~125 K.

#### 3.2.2. Field dependence of magnetization

Magnetization was recorded as a function of applied field (M(H)) at 5 K and 300 K. Fig. 9 shows the results of M(H) measurements on the films: as-deposited and 673 K annealed for 2 h and 8 h respectively. It is evident from the figure that there was no notable difference between 5 K and 300 K measurements, as both temperatures are well below the known Curie temperature of bulk Ni. Further, no hysteresis was observed. Instead a linear dependence of magnetization on field in the high-field range, as expected of a superparamagnetic material [51] was obtained. The phase essentially consisted of Ni particles as evidenced by other studies described in this paper. Superparamagnetism is a property associated with ferromagnetic nanoparticles. It occurs in nanoparticles which are composed of a single magnetic domain. Superparamagnetism in Ni fine particles has been reported earlier [52]. The magnetization of the nanoparticles is thus considered as a single giant magnetic moment, arising from individual magnetic moments. When an external magnetic field is applied to an assembly of superparamagnetic nanoparticles, their magnetic moments tend to align along the applied field, leading to a net magnetization (reversible S-shaped increasing function seen in

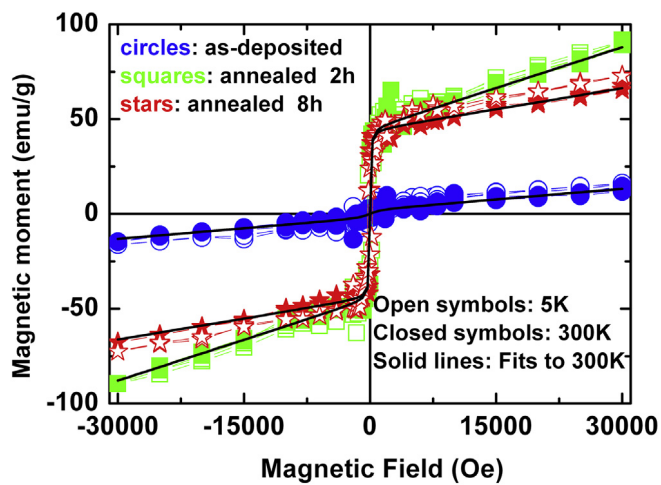

**Fig. 9.** Magnetization of as-deposited and annealed films shown as a function of applied magnetic field, at 5 K and 300 K.

figure). It is worth pointing out that prolonged annealing at 673 K brought down the linear component of M(H), which indicates an enhanced tendency of saturation expected from a ferromagnetic material. The magnetization curve M(H) at 300 K was fitted with a modified Langevin function, $M(H) = M_s\, L(x) + \chi_f\, H$, where $L(x) = [\coth(x)-(1/x)]$ with $x = \mu H/kT$, $\mu$ is the average magnetic moment of the magnetic particle/entity, $M_s$ the saturated magnetization and $\chi_f$ is the high field susceptibility of the sample showing a linear dependence on $H$. Fits are shown as black solid lines in Fig. 9. From the fit, saturation magnetization of annealed samples was extracted to be ~45 emu/g, which translates to 0.47 $\mu_B$/Ni-atom, in agreement with PNR estimates.

### 3.3. Molecular dynamics simulations

The nearest neighbor analyses of MD simulations for both bulk and slab of $Ni_{10}Zr_7$ are presented in Table 1, which lists average number of Ni and Zr atoms around a Ni (Zr) atom within a cut-off distance of 7.5 Å before and after annealing at 673 K. The results obtained on annealing at 873 K do not differ much from what has been presented here. As can be seen in the table, the average number of Ni atoms around a Ni atom increases in the bulk (surface slab) from 66.90 (63.36) to 69.40 (66.95) on annealing, indicating clustering tendencies present in the system. This is also evidenced by a reduction in the average number of Zr atoms around a Ni atom from 50.60 (48.30) to 48.36 (46.80). Similarly, the average number of Zr atoms around a Zr atom increases from 45.10 (42.59) to 47.60 (45.10), while the average number of Ni atoms around Zr decreases from 72.10 (70.81) to 68.84 (67.10) lending credence to the segregation observed in both bulk and the surface.

A pair correlation function gives an idea of variation of number of atoms from a reference atom (distribution) with distance. It also gives information about the structure (crystalline or amorphous) of the material. The pair correlations (g(r)) for Ni–Ni and Zr–Zr pairs before and after simulated annealing are plotted in Fig. 10 and clearly show an increase after annealing. From the above results, it may be deduced that in $Ni_{10}Zr_7$ there is a tendency of Ni and Zr atoms to form Ni-rich and Zr-rich regions.

Migration energy can be defined as the minimum energy required to initiate the diffusion process or can be a measure of a diffusional barrier. It can also be thought of as an energy barrier between two local minima of the potential. The migration energies of Ni and Zr in bulk $Ni_{10}Zr_7$ have also been calculated. These results are shown in Fig. 11, where one can see that the migration energy of Ni (1.031 eV) is much less than that of Zr (2.01 eV), which clearly indicates the easier diffusion of Ni atoms as compared to that of Zr atoms.

### 4. Summary and conclusions

This study demonstrates the creation of a crystalline alloy in a binary system with a tendency to amorphize. Nanocrystalline $Ni_{10}Zr_7$ alloy films were successfully grown by co-sputtering after tuning deposition parameters in an indigenous D.C./R.F. magnetron sputtering system. The barrier to crystalline structure formation usually observed in the Ni–Zr system was thus overcome. The co-sputtered film was formed in single phase and was also characterized to be homogeneous and uniform. This study also elaborates on the thermal evolution of the crystalline phase. The nanocrystalline alloy films were found to be stable with ex situ annealing at 473 K but annealing at 673 K caused the segregation of crystalline Ni towards the air-film interface from a background of amorphous Ni + Zr phases. Several surface and interface sensitive techniques were employed to examine the chemical and magnetic structure of the as-deposited and annealed films in detail to understand the dynamics of phase separation, and their results were corroborative. The phase separation after annealing could be adequately explained though free energy considerations, supported by MD simulations. With increasing annealing time, the growth of crystalline Ni grains was observed. The as-deposited alloy film was found to be almost non-magnetic but after annealing, magnetic measurements brought out features typical of a superparamagnetic phase.

**Table 1**
Average co-ordination numbers of Ni and Zr atoms before and after annealing in bulk and surface slab.

| Type of Pair | Bulk | | Slab | |
|---|---|---|---|---|
| | Before | After | Before | After |
| Ni–Ni | 66.90 | 69.40 | 63.36 | 66.95 |
| Ni–Zr | 50.60 | 48.36 | 48.30 | 46.80 |
| Zr–Ni | 72.10 | 68.84 | 70.81 | 67.10 |
| Zr–Zr | 45.10 | 47.60 | 42.59 | 45.10 |

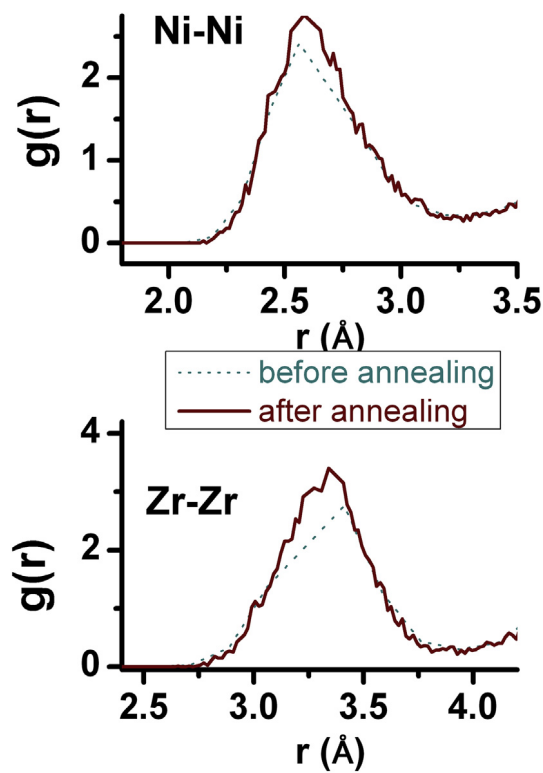

**Fig. 10.** Simulated pair correlation functions for Ni–Ni and Zr–Zr in as-deposited alloy.

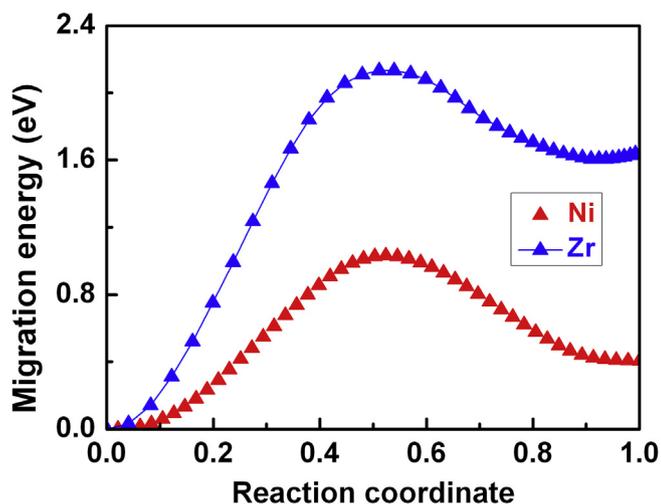

**Fig. 11.** Simulated migration energies for Ni and Zr in as-deposited alloy.

## Acknowledgments

The author wishes to thank Dr. (Mrs) Seemita Banerjee for useful discussions regarding applications of Ni–Zr alloys and Mr. Swapan Jana for assistance during film deposition.